\documentclass[twocolumn,aps,prl,showpacs]{revtex4}

\usepackage{epsfig}
\usepackage{graphics}
\usepackage{soul}
\usepackage{amsmath}

\begin{document}
\title{Faithful hierarchy of genuine n-photon quantum non-Gaussian light}
\author{Luk\' a\v s Lachman \footnote{lachman@optics.upol.cz}}
\thanks{These authors contributed equally to this work.}
\author{Ivo Straka}
\thanks{These authors contributed equally to this work.}
\author{Josef Hlou\v sek}
\author{Miroslav Je\v zek}
\author{Radim Filip}

\affiliation{Department of Optics, Faculty of Science, Palack\' y University,\\
17. listopadu 1192/12,  771~46 Olomouc, \\ Czech Republic}
\begin{abstract}
Light is an essential tool for connections between quantum devices and for diagnostic of processes in quantum technology. Both applications deal with advanced nonclassical states beyond Gaussian coherent and squeezed states. Current development requires a loss-tolerant diagnostic of such nonclassical aspects. We propose and experimentally verify a faithful hierarchy of genuine $n$-photon quantum non-Gaussian light. We conclusively witnessed 3-photon quantum non-Gaussian light in the experiment. Measured data demonstrates a direct applicability of the hierarchy for a large class of real states.
\end{abstract}
\pacs{42.50.Xa, 42.50.Ar, 42.50.Dv}
\maketitle

\begin{figure}
\centerline {\includegraphics[width=0.95\linewidth]{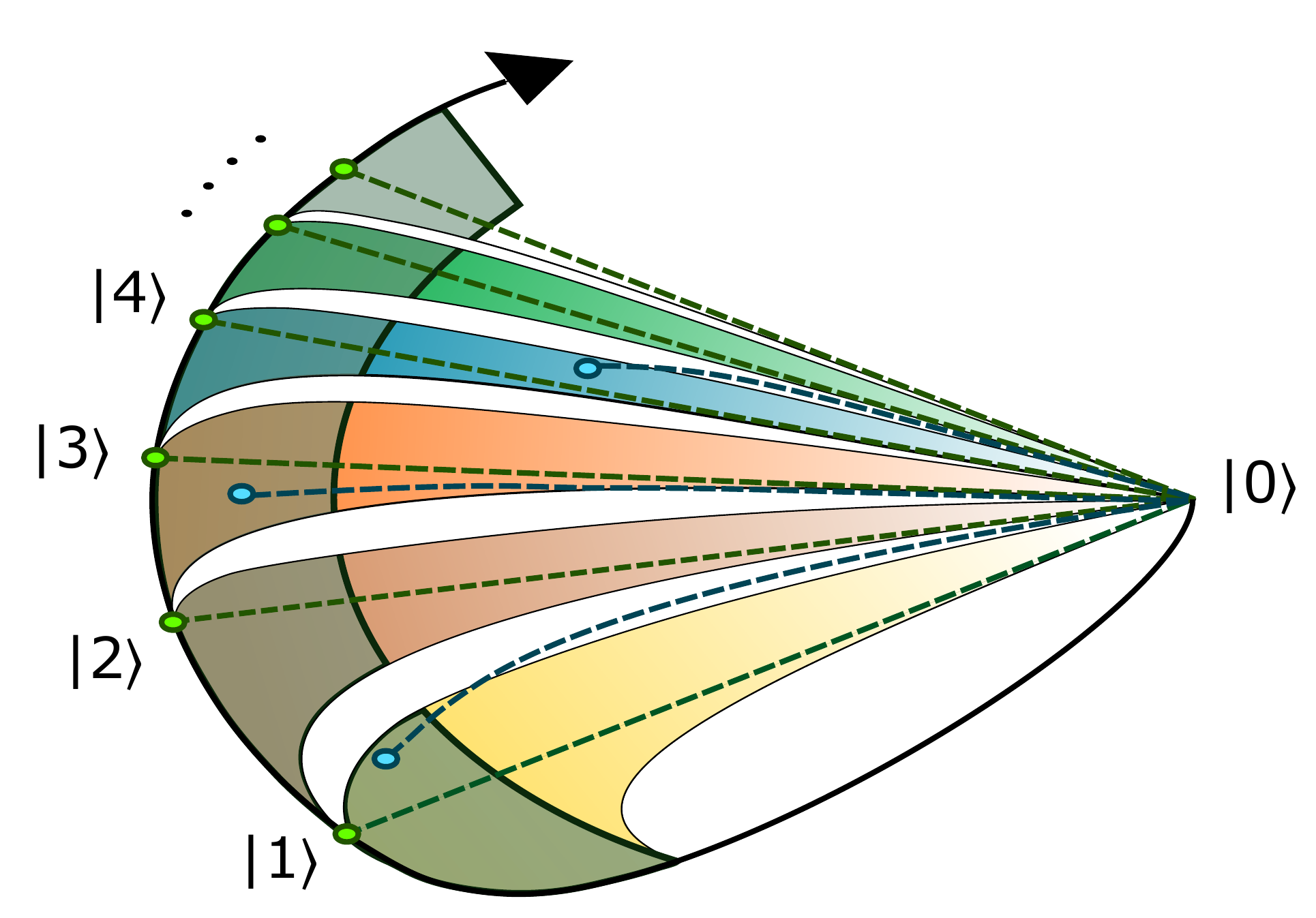}}
\caption{A visual presentation of the hierarchy of genuine quantum non-Gaussian states approaching ideal Fock states of light. The white regions stand for mixtures of Gaussian states (squeezed coherent states). All colored regions represent states beyond those mixtures. Each color corresponds to a new quantum feature attached to highly nonclassical states such as Fock states $\vert n \rangle$ (green points). The hierarchy of such features classifies multiphoton light exhibiting quantum non-Gaussianity. Advantageously, these features are more robust against attenuation than negativity of the Wigner function (opaque gray regions). The quantum non-Gaussianity of ideal Fock states manifests absolute robustness against losses. Realistic states approaching the Fock states (blue points) can lose the genuine quantum non-Gaussianity when they are affected by losses. The blue and green dashed lines represent the influence of attenuation on states exhibiting genuine $n$-photon quantum non-Gaussianity.}
\label{fig1}
\end{figure}

Individual photons as bosonic elementary particles have been subjects of a detailed quantum analysis already for many decades. It is intensified now due to their importance for quantum technology. First, a single photon antibunching was measured as incompatible with classical coherence theory \cite{mandel, grangier}. It was the first proof of nonclassical light. This measurement became canonical for single photon sources \cite{mandel,grangier,PDC,QD,yamamoto}. After many years, broadband homodyne detection allowed indirect estimation of their continuous variable nonclassical features \citep{lvovskyHM,lvovsky, grangierHD, furusawa, smith, laiho}. Their visualisation in the phase space of continuous amplitude of the electric field by a Wigner quasiprobability distribution shows multiple negative concentric annuli for Fock states of light \cite{Schleich}. The Wigner function is used to distinguish different Fock states of light, however, without any proof yet that they really form a {\em faithful} hierarchy. A faithful hierarchy of $n$-photon quantum non-Gaussianity would reliably recognize that, for a given order $n$, an observed state is statistically incompatible with any mixture of Fock-state superpositions up to $|n-1\rangle$ modified by an arbitrary Gaussian phase-space transformation \citep{furusawa, fock2, smithf3}. The hierarchy is schematically presented in Fig.~\ref{fig1}. Unfortunately, such a faithful hierarchy based on the negative parts of the Wigner function has not been discovered yet and it would be anyway applicable only if overall losses were below fifty percent \citep{ivoAtt}. Since a large variety of experimental platforms emitting or transmitting light does not suppress the losses so much, a lack of theoretical tools witnessing genuine $n$-photon quantum non-Gaussianity limits optical diagnostic of quantum processes in matter, current fast development of multiphoton sources and their applications in quantum technology.

A large gap between basic nonclassical light and light with a negative Wigner function was partially covered when a loss-tolerant direct measurement of single-photon quantum non-Gaussianity was proposed and immediately experimentally tested \citep{mista,jezek}. Advantageously, these criteria use only basic multiphoton correlation measurements, commonly applied to verify nonclassicality. The quantum non-Gaussianity criteria conclusively prove that a quantum state of light is not compatible with any mixture of Gaussian states, even beyond fifty percent of loss \citep{ivoAtt}. In difference to the tests of nonclassicality, such tests of quantum non-Gaussianity can already recognize a much narrower set of states, approaching closer to ideal single photon states.
That property of single photon states has already been proposed to be applicable as a security indicator of single-photon quantum key distribution \citep{lasota} and as a probe of quantum photon-phonon-photon transfer \citep{Rakhubovsky}. In both cases, it was proven that a test of nonclassicality is not sufficient and it can be misleading. Recently, criteria of quantum non-Gaussianity for multiphoton light have been proposed and measured despite very large optical loss \citep{ivo}. Meanwhile, quantum non-Gaussianity criteria have been developed for other types of states \citep{paris, genoni,vogel, happ}. Recent mathematical treatment of quantum non-Gaussianity led to a formulation of a resource theory \citep{resourceTh1,resourceTh2}.

The extension to multiphoton light allows wider applications in diagnostic of quantum processes, but the criteria \cite{ivo} do not still form a faithful hierarchy of quantum properties and  therefore, genuine $n$-photon quantum non-Gaussian state can not be directly witnessed under large optical loss. The discovery of the hierarchy is currently crucial for ongoing exploration of light emitted by higher order nonlinear processes \citep{triplets1,triplets2} and for current development of multiphoton sources \citep{fock2,weihs}. In this Letter, we derive the faithful hierarchy of sufficient conditions for genuine $n$-photon quantum non-Gaussian states and, simultaneously, we experimentally verify the hierarchy by measuring multiphoton light up to three heralded photons under 6.5 dB of optical loss. Under such loss, a negative Wigner function cannot be observed. Our criteria can conclusively confirm that the observed genuine $n$-photon quantum non-Gaussian statistics is beyond statistics produced by any mixture of superposition of $n-1$ photons possibly modified by any Gaussian transformation. 

A pure state $\vert \psi \rangle$ exhibits genuine $n$-photon quantum non-Gaussianity if it can not be expressed as
\begin{equation}
\vert \psi \rangle \neq  S(\beta)D(\alpha)\vert \widetilde{\psi}_{n-1} \rangle,
\label{nnonG}
\end{equation}
where the core state $\vert \widetilde{\psi}_{n-1} \rangle =\sum_{m=0}^{n-1}\tilde{c}_m|m\rangle$ represents any superposition of the Fock states $\vert 0 \rangle,..., \vert n-1 \rangle$ that can be affected by displacement $D(\alpha)=\exp (\alpha a^{\dagger}-\alpha^* a)$ or by squeezing $S(\beta)=\exp \left[-\beta \left(a^{\dagger}\right)^2+\beta^* a^2 \right]$ operation. The transformation $S(\beta)D(\alpha)$ can add only a Gaussian envelope to the core state $|\widetilde{\psi}_{n-1}\rangle$ \cite{menzies}. The envelope changes photon statistics; however, it only scales the shape of the Wigner function representing the state in the phase space. A single-mode mixed state $\rho$ shows the $n$th-order property if it cannot be identified with any statistical mixture of the right side in inequality (\ref{nnonG}) randomized over the complex parameters $\alpha,\ \beta$ and $\tilde{c}_m$. 
That introduces a hierarchy of genuine quantum non-Gaussian attributes labeled by an index $n$. Obviously, each ideal Fock state $\vert n \rangle$ possesses the attribute of the order $n$ which any lower Fock state cannot achieve through any Gaussian transformation. Also, any state $\vert n \rangle$ attains the attributes of order $m <n$.
The lowest (first) order of the hierarchy represents quantum non-Gaussianity proposed and measured in Refs.~\citep{mista, jezek}. The second order means that observed photon statistics are not compatible with any mixture of states $S(\beta)D(\alpha)\left(\tilde{c}_0|0\rangle+\tilde{c}_1|1\rangle\right)$ for any complex $\alpha$, $\beta$, $\tilde{c}_0$ and $\tilde{c}_1$ satisfying $|\tilde{c}_0|^2+|\tilde{c}_1|^2=1$.  
In this case, the Gaussian transformation $S(\beta)D(\alpha)$ increases the number of photons beyond one, but it does not extend the genuine non-Gaussian attribute to $n=2$, which is typical for the Fock state $\vert 2 \rangle$.

\begin{figure*}
\centerline {\includegraphics[width=18cm]{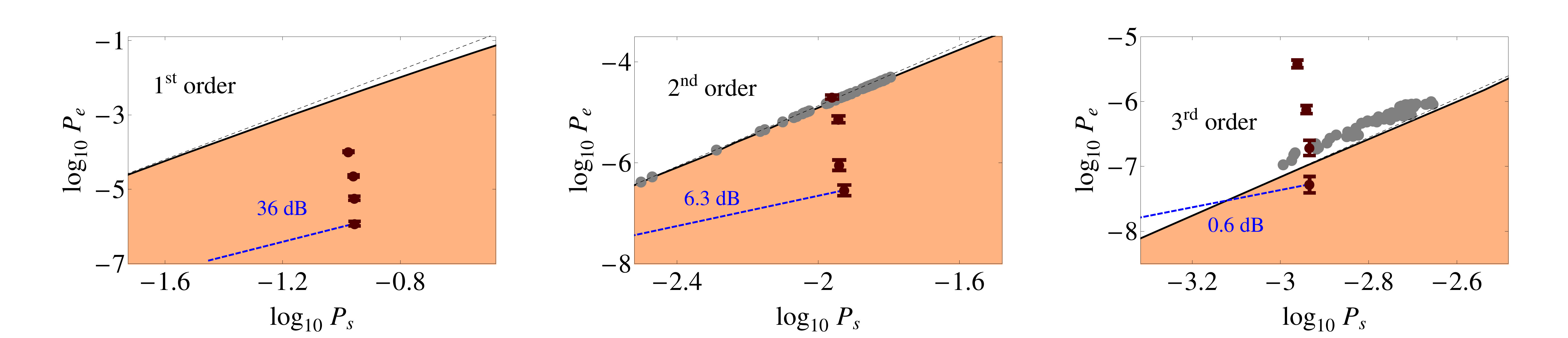}}
\caption{The faithful hierarchy witnessing the genuine $n$-photon quantum non-Gaussianity up to order three and its experimental verification. The quantum non-Gaussianity is recognized in the orange regions. The approximate solutions of the thresholds for $n=1,\ 2,\ 3$ are plotted by dashed lines. The reliability of the thresholds is demonstrated by the results of a Monte Carlo simulation. The gray points represent fifty samples generated in the simulation that were closest to the threshold. The total number of runs in the simulation was $10^6$ (2nd order) and $10^8$ (3rd order). The black points correspond to the experimental data. The shifting of the points along the vertical axes corresponds to deterioration of the emitted light by background Poissonian noise. The slight movement of the points in the horizontal axis is caused by experimental imperfections resulting in noise leakage into the heralding arm.  The mean number of photons of the background noise registered in a detection window is $\bar{n}=0, 4 \times 10^{-5}, 2 \times 10^{-4}, 10^{-3}$ (from the lower points to the upper points) for each measurement. Error bars represent statistical error of the number of detected coincidence events. The effects of optical loss on the experimental data are illustrated by the blue dashed lines. Attenuated states would follow these lines until they would cross the thresholds. The attached values represent the maximum attenuation, under which genuine quantum non-Gaussianity remains observable.}
\label{expSet}
\end{figure*}

The criteria will be derived ab initio without any assumptions about the inspected states of light. Thus, they can be applied to any states with any mean number of photons. As such, the criteria depend only on the formulation of the detection process. 
The witnessing of genuine quantum non-Gaussianity is provided by a balanced multichannel detector. Incoming light is evenly split to $n+1$ single-photon avalanche diodes (SPADs) that only distinguish signal from vacuum. The genuine $n$-photon property is detected when the probability of simultaneous detections on all $n+1$ SPADs (error) is suppressed sufficiently relative to the probability of simultaneous $n$ detections (success). Let us choose a set of $n$ detectors and define the probability of their simultaneous detection by $P_s$ and the probability of all $n+1$ detectors clicking by $P_e$. In this case, $P_s$ refers to the probability of an expected success event, when light contains at least $n$ photons, and $P_e$ quantifies the probability of an unwanted error event, when light contains at least $n+1$ photons. A linear combination of both probabilities
\begin{equation}
F_{a,n}(\rho)=P_s+a P_e,
\label{fa}
\end{equation}
where $a$ is a free parameter, identifies $n$-photon genuine quantum non-Gaussianity if
\begin{equation}
 \exists a: P_s+a P_e>F_n(a),
 \end{equation}
where $F_n(a)$ is a threshold function that is determined from optimizing the functional $F_{a,n}(\rho)$ over mixtures of states given by the right side of (\ref{nnonG}) with the order $n$. The subscript $n$ also denotes the number of SPADs required for a success event. Note that the condition can be also formulated so that the number of detectors identifying success is greater than the order of the witnessed property. In that case, the criterion applied on a Fock state $\vert n \rangle$ reveals its attribute with a lower order than $n$. Because the functional $F_{a,n}(\rho)$ is linear with respect to quantum states, the optimum is obtained as a pure state $S(\beta)D(\alpha)\vert \widetilde{\psi}_{n-1} \rangle$ where $\vert \widetilde{\psi}_{n-1} \rangle=\sum_{k=0}^{n-1} \tilde{c}_k \vert k \rangle$. The state is formally expressed by $2 n+4$ parameters which hold normalization. Since two states with different global phases are identical, the considered state is determined by $2(n+1)$ unique parameters. The task is finding an optimum over these parameters. This can be performed only numerically. The Supplementary Material provides a detail description of algorithm which searches for the maximum. The derived thresholds are depicted in Fig.~\ref{expSet} for layouts with three and four SPADs. The algorithm has to incorporate extensive formulas that express general parametrization of success and error probabilities \citep{kral}. However, assuming that the inspected states have a strongly suppressed probability of error $P_e$, as is typical for high-quality multiphoton states, the threshold can obtain approximate forms
\begin{equation}
P_e\approx\frac{(1+n)^{2n}(2+n)^2(1+n)!P_s^3}{18n^2(n!)^3}.
\label{thresApp}
\end{equation}
These approximations are applicable as a basic witness, however, they are below the real thresholds. They have to be carefully used if data are very close to them, surpassing them too tightly can lead to a false positive. Thus, the Supplementary Material provides a derivation of more accurate approximations that can be applied on a larger set of states.
Also, usefulness of our approach is presented in the Supplementary Material. It is demonstrated there that our method can identify the presence of $n$-photon genuine quantum non-Gaussianity; even among states that share almost identical photon statistics.

Let us note that although the thresholds were derived from the assumption of single-mode states, they can be applied to states occupying multiple modes. This is also the case in the presented experimental proof. The genuine $n$-photon quantum non-Gaussianity of multi-mode states means the higher photon contributions are produced neither by squeezing nor by displacement of a multi-mode core state that shows a truncated photon distribution. Since the exact definition of that property is technical in the multi-mode case, it is presented in the Supplementary Material along with the details of a Monte Carlo simulation indicating the thresholds do not get stricter for multi-mode states.

{\em Experimental setup.---} To experimentally witness genuine $n$-photon quantum non-Gaussianity, a superior multiphoton source is required. In this regard, there has been development recently reported in the works \citep{goetzinger,lodahlSP,senellart}. We generated statistics with controllable multiphoton content from a well-established photon source based on multiple high-quality single photons triggered to suppress random noise. We employed continuous-wave spontaneous parametric down-conversion (SPDC) to generate sequences of $n$ heralded single photons that were collectively measured on a multichannel detector as depicted in Fig.~\ref{setup}. See the Supplementary Materials for further details about the source. In addition to this signal, we added extra Poissonian background noise from a laser diode to explore the sensitivity of genuine $n$-photon quantum non-Gaussianity to multiphoton content. We detected photons from all triggered time modes collectively, considering the overall statistics.

\begin{figure}
\centering
\includegraphics[width=.4\textwidth]{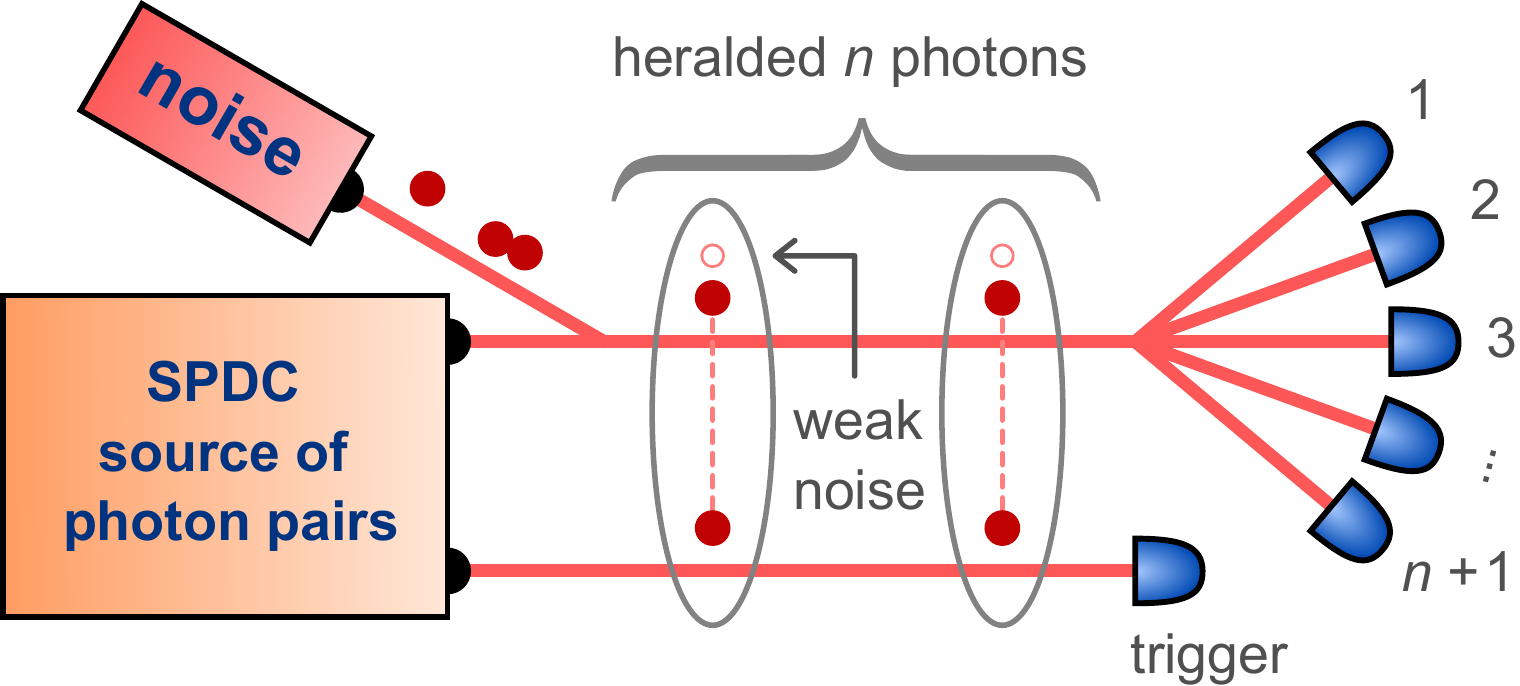}
\caption{Schematics of the experiment. A number of down-converted heralded photons with weak multiphoton contributions are incident on a balanced multi-channel detector consisting of single-photon avalanche diodes (SPAD). The multiphoton contribution consists of multi-pair generation and, primarily, of Poissonian noise added by coupling a laser diode to the signal. Only time windows conditioned by a trigger detection are considered, and $n$ successive time windows are merged together.}
\label{setup}
\end{figure}

\begin{figure}
\centerline {\includegraphics[width=0.85\linewidth]{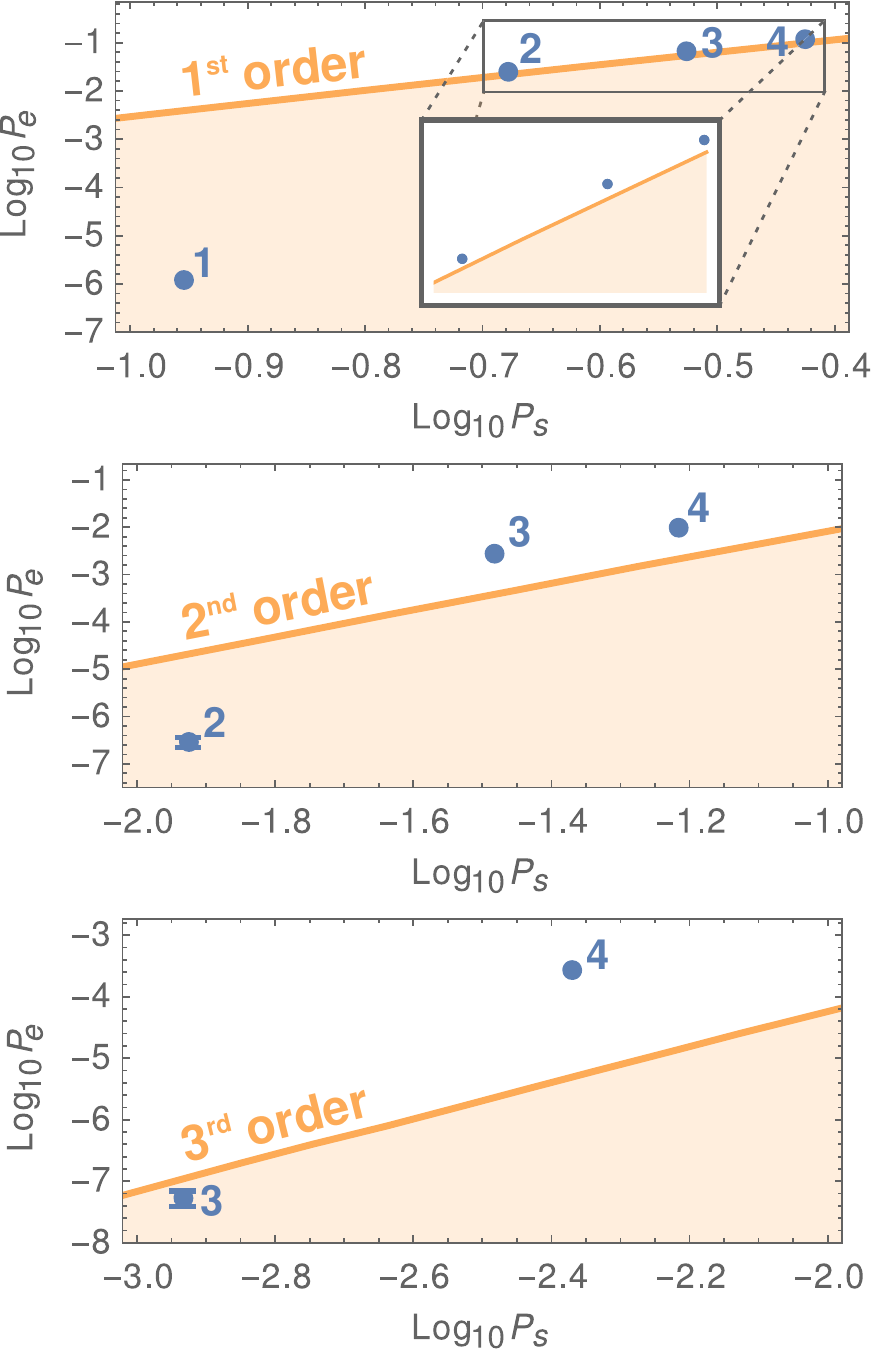}}
\caption{For each order of genuine quantum non-Gaussianity  (orange), various heralded numbers of photons are plotted. The blue points are experimental results with the respective numbers of merged photons attached. Error bars are shown only when the uncertainty is comparable to the point size. The only states passing the respective criteria are those with a matching number of photons -- those are also shown in Fig. \ref{expSet} with expected attenuation paths. The differences in scale of each graph are caused by varying the number of detector channels and scaling of the $n$-photon probabilities with increasing $n$.}
\label{dataOrders}
\end{figure}

The detector was implemented by a balanced network of half-wave plates and polarizing beam splitters with a silicon SPAD in each arm. The temporal resolution of the SPADs was safely covered by the coincidence windows. We recorded coincidence events between individual SPADs and obtained results presented in Figs. \ref{expSet} and \ref{dataOrders}.
The estimated efficiencies of the SPADs were between 50 and 65 $\%$. The differences between overall efficiencies of each detector arm were compensated by adjusting the splitting ratios to balance detection rates among all SPADs. The result is equivalent to a balanced detector with a fixed overall efficiency. By virtue of definition of the genuine quantum non-Gaussianity, the finite efficiency of the detector -- contributing only vacuum -- cannot cause false witnessing. Therefore any such witnessed quantum state is indeed genuinely quantum non-Gaussian.

{\em Results and analysis.---} The data exhibit genuine quantum features up to order three ($n=3$).
This was achieved by minimizing SPDC gain and the time window for coincidence detection, because $P_e$ grows linearly with both parameters. The time window is limited by the temporal resolution of the detectors, while the gain can be lowered arbitrarily at the cost of reducing generation rate. The experimental limit of our demonstration was the measurement time needed to acquire statistically significant results for $P_e$. The scaling is very fast: while we needed only 16 hours to obtain the results for $n=3$, several months would be needed for $n=4$. The main factor is that lowering SPDC gain simultaneously decreases the portion of error events and generation rate. A low event rate limits similarly also other experiments demonstrating a negative Wigner function of a three-photon state \cite{furusawa, cooper, silberhorn50}. To maintain a sufficiently low error rate with an increased gain, the detection time window would have to be reduced. This parameter is limited by the temporal resolution of SPADs and could be augmented by using detectors optimized for low jitter. Optical loss in both arms of the source, including detection loss, is also a factor, which depends on coupling efficiency as well as detector efficiency. The final limiting factor are the background dark counts, which become relevant in the extremal case of a very low gain and long measurement. Overall, the detection precision, signal-to-noise ratio and efficiency represent the main factors in the presented type of measurement.
Fig.~\ref{expSet} shows that robustness against losses and noise rapidly decreases with higher order. This is a consequence of decreasing the maximum gain allowable for higher $n$. Our data were all measured with the same gain, which means the individual statistics of all constituent heralded events are the same. The relation between the number of heralded events and successful witnessing of genuine $n$-photon quantum non-Gaussianity is presented in Fig.~\ref{dataOrders}. If the number of heralded events exceeds the order of the witnessing criterion, the detection of that property in our measurement fails. However, the property of the order $n$ is implied by its positive recognition for any order greater than $n$. Furthermore, the criteria for the order $n$ can be reformulated for any higher number of detector channels than $n+1$. That is because the functional (\ref{fa}) can be maximized over any set of states defined by the right side of inequality (\ref{nnonG}), even if the order of the state and the number of detector channels do not match. When the order of the criterion is greater than the number of heralded events, the property cannot be detected solely because of its definition. These cases are however not depicted in the Fig.~\ref{dataOrders} due to the scale of confidence intervals of relevant error and success probabilities.

{\em Outlook.---}
The presented hierarchy of genuine quantum non-Gaussianity for the states approaching Fock states of light and its experimental verification can be applied to a class of new multiphoton experiments \citep{goetzinger,lodahlSP,senellart} and to observe quantum non-Gaussianity of first photonic triplets \citep{weihs} from cubic nonlinear materials \citep{birnbaum, peyronel, chang, firstenberg, javadi, snijders, sipahigil, bhaskar}. As such, it can stimulate further experimental research in this pioneering direction of quantum technology with multi-photon states of light. Ab initio approach to the hierarchy allows further extensions towards different quantum non-Gaussian states and its multi-mode versions used for both fundamental tests \cite{sychev,Laurat} as well for applications in quantum technology with light \cite{ulanov, makino}. Because light is dominantly used for read-out from atomic and solid state systems, this methodology can be used and also extended to evaluate quantum non-Gaussianity of, for example, already developed atomic-ensemble memories \citep{chaneliere,eisaman,choi, distante} and new single-phonon mechanical oscillators \citep{shong}. 

\section*{Acknowledgement}

L.L. and R.F. acknowledge the support of the project
GB14-36681G of the Czech Science Foundation, the support of the project TheBlinQC of QuantERA within ERA-NET Cofund in Quantum Technologies and 
IGA-Prf-2018-010.

\section{Supplementary information}
\begin{figure}
\centerline {\includegraphics[width=0.95\linewidth]{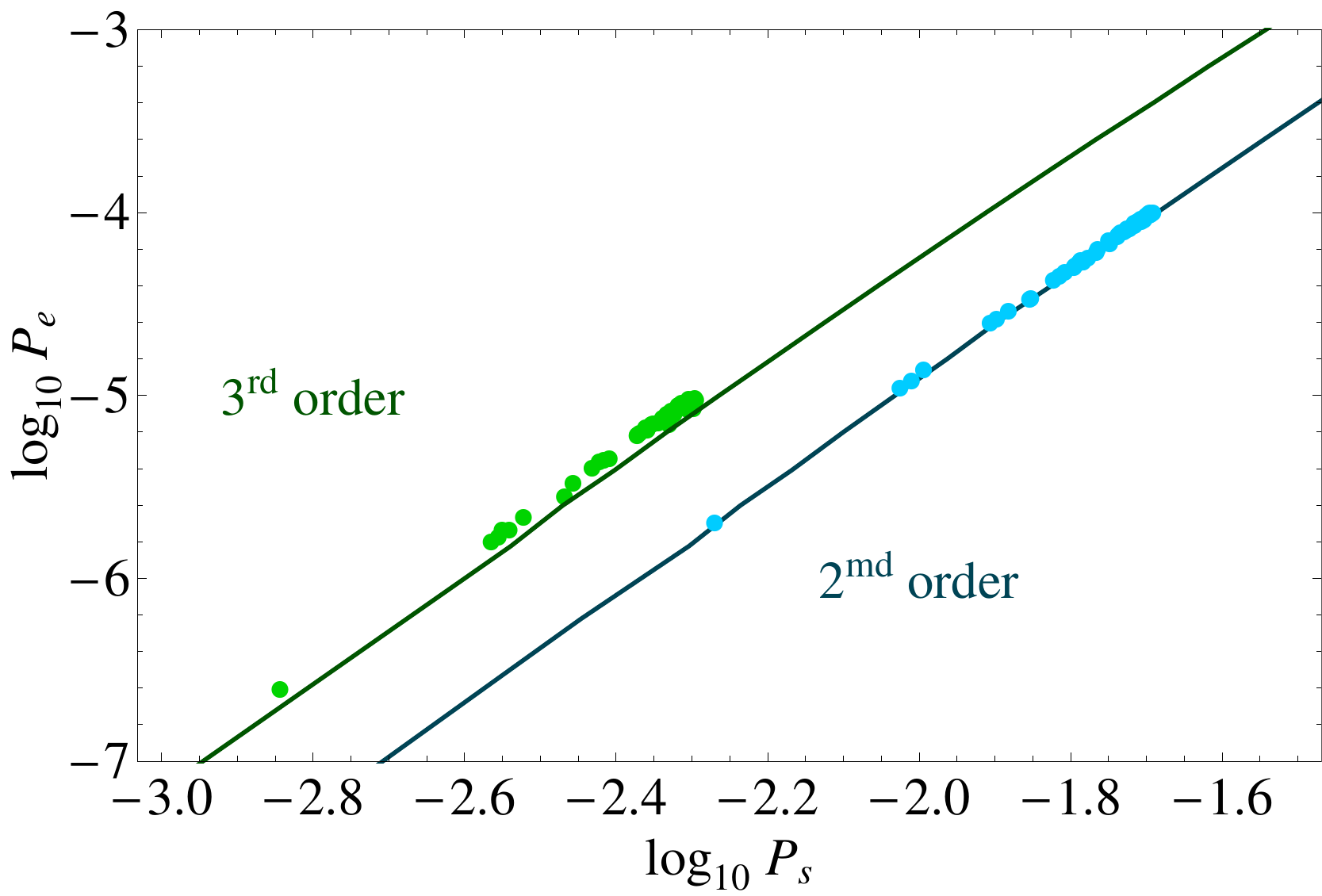}}
\caption{The results of Monte-Carlo simulation verifying covering two mode states that are produced by squeezing and displacement of two mode core state $\vert \widetilde{\psi}_{n-1} \rangle$. The plot presents fifty points generated closest to the thresholds of second (blue points) and third order (green points). The number of runs of the simulation was set to $10^8$ in both cases.}
\label{fig1SM}
\end{figure}

The formula (1) in the main part of the manuscript defines a hierarchy of quantum non-Gaussian properties only for single mode states.
The genuine $n$-photon quantum non-Gaussianity recognize statistically significant highly nonclassical features of light and thus it can be extended even for multi-mode states. Let $M$ denotes number of considered modes. Similarly as in the single mode case, the extended definition involves a statistically truncated core state $\vert \widetilde{\psi}_{n-1} \rangle$ that exhibits
\begin{equation}
\langle m_1 \vert \otimes ... \otimes \langle m_M \vert \widetilde{\psi}_{n-1} \rangle \neq 0
\end{equation}
only if $\sum_{i=1}^M m_i < n$, where $\langle m_i \vert$ is a Fock state $m$ occupying the $i$th mode. The constrain guarantees the state does not produce $n$ or more than $n$ photons. In a single mode case, the core state can be expressed as $\vert \widetilde{\psi}_{n-1} \rangle =\sum_{k=0}^{n-1} c_k \vert k \rangle$. Two modes core states correspond to $\vert \widetilde{\psi}_{n-1} \rangle= \sum_{k=0}^{n-1} \sum_{l=0}^{n-k-1}C_{k,l} \vert k, l \rangle$ with arbitrary coefficients $C_{k,l}$. The higher photon contributions of the refused states can be generated only by squeezing or displacement influencing all modes occupied by $\vert \widetilde{\psi}_{n-1} \rangle$. Let $S_i(\beta_i)$ be squeezing and $D_i(\alpha_i)$ be displacement affecting mode $i$, where $\beta_i$ and $\alpha_i$ are parameters determining the operators. A pure multi-mode state $\vert \psi \rangle$ exhibits genuine $n$-photon quantum non-Gaussianity if
\begin{equation}
\vert \psi \rangle \neq S_M(\boldsymbol{\beta})D_M(\boldsymbol{\alpha}) \vert \tilde{\psi}_{n-1} \rangle,
\label{mnonG}
\end{equation}
where $\boldsymbol{\alpha}$ and $\boldsymbol{\beta}$ are vectors $\boldsymbol{\alpha}=(\alpha_1,...,\alpha_M)$, $\boldsymbol{\beta}=(\beta_1,...\beta_M)$ and $S_M(\boldsymbol{\beta})$, $D_M(\boldsymbol{\alpha})$ are defined as tensor products
\begin{eqnarray}
S_M(\boldsymbol{\beta})&=&\Pi_{i=1}^M \otimes S_i(\beta_i) \nonumber \\
D_M(\boldsymbol{\alpha})&=&\Pi_{i=1}^M \otimes D_i(\alpha_i).
\end{eqnarray}
The genuine $n$-photon quantum non-Gaussianity also refuses all statistical mixtures of right side of inequality (\ref{mnonG}).
It remains to check that function $F_n(a)$ covers all states parametrized by the right side of inequality (\ref{mnonG}). A partial proof can be find for states with low probability of error. The only general approach is Monte-Carlo simulation which, however, requires further software development to be performed due to a large number of parameters determining a general multi-mode $n$-photon Gaussian state. The simulation for two mode states was carried out only with real coefficient of the core state $\vert \widetilde{\psi}_{n-1} \rangle$ and squeezing orthogonal to displacement. That decreases a number of parameters over which the optimum is searched to $\frac{1}{2}n(n+1)+1$. The results are presented in Fig.~\ref{fig1SM}  for the second and third order of the criteria.

\begin{figure*}
\centerline {\includegraphics[width=18cm]{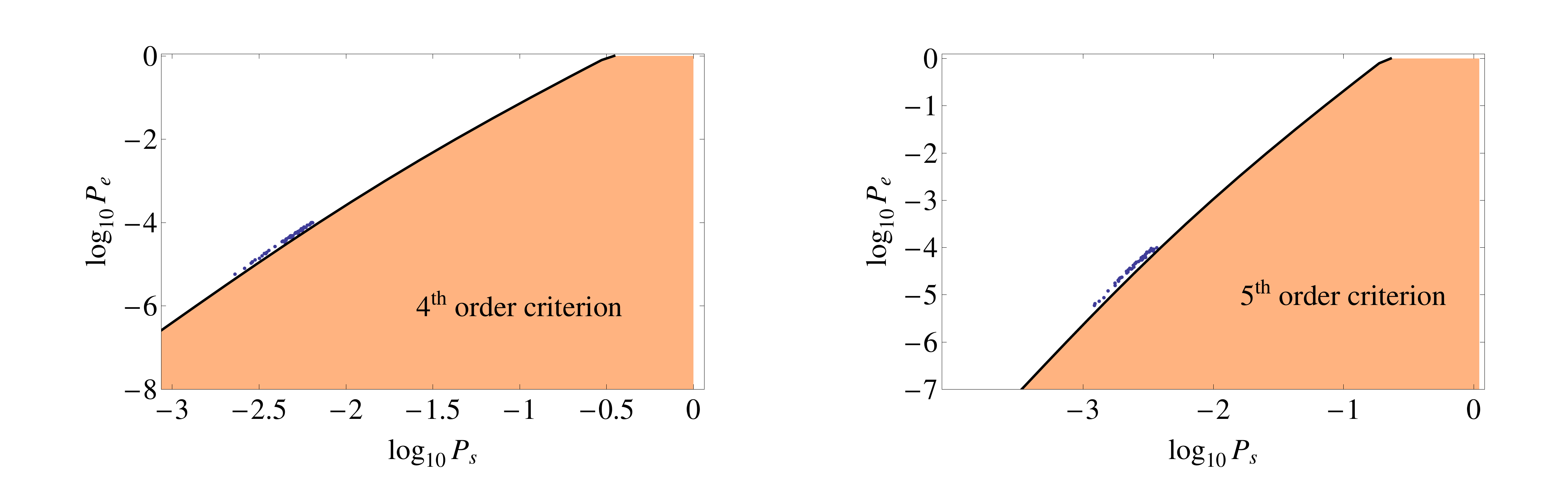}}
\caption{Monte-Carlo verification of hierarchy of approximate conditions (\ref{appTh}) for 4th and 5th order. The blue points represent fifty results that were generated closest to the thresholds. The total number of attempts in the both simulations was $10^8$. The simulation was restricted to states with $P_e<10^{-4}$, where the approximations are expected to fix the real thresholds very tightly.}
\label{MCsimSM}
\end{figure*}

The limit of states with low probability of error involves states with small squeezing and small displacement. Then, the error and success probabilities can be approximated by formulas
\begin{eqnarray}
P_s& \approx &C_{n,n}P_{n}+C_{n,n+1}P_{n+1}\nonumber \\
P_e& \approx &C_{n+1,n+1} P_{n+1}+C_{n+1,n+2}P_{n+2}+ \nonumber \\
&+&C_{n+1,n+3}P_{n+3},
\label{pspe}
\end{eqnarray}
where $n$ represents a number of simultaneous clicks identifying success, $P_k$ is a probability of having $k$ photons in the input of the test and $C_{n m}$ denotes probability that $m$ photons cause simultaneous clicks of $n$ detectors. The error probability requires to be expanded by more members because the parameters of the state can be set such that $P_{n+1} \sim P_{n+2} \ll 1$. The matrix $C_{n m}$ holds \citep{smRef}
\begin{equation}
C_{n m}=1+\sum_{k=1}^n {n \choose k} (-1)^k \left(1-\frac{k}{N}\right)^m,
\end{equation}
where $N$ is total number of detector. 
The state $\vert \widetilde{\psi}_{n-1} \rangle = \sum_{k=0}^{n-1} c_k \vert k \rangle$ on which squeezing and displacement is performed is assumed to be identical with Fock state $\vert n-1 \rangle$ in the approximation. Also, the conjecture that the threshold is derived from single mode states is partially checked in this regime because only single mode Gaussian states can exhibit $P_{n+1} \sim P_{n+2}$.
In this regime, the solution of equation (4) in the main text gains
\begin{eqnarray}
A(V,\vert n-1 \rangle) \approx 2 \beta+\frac{4}{3}(3+2n+n^2)\beta ^2,
 \end{eqnarray}
 where $0<\beta \ll 1$.
Inserting it into formulas (\ref{pspe}) results in approximates
 \begin{eqnarray}
P_e &\approx &\frac{n n!(n+2)^2}{55296 (n+1)^{n-1}}\beta ^3\left[384+\beta(896+307n+99n^2)\right]\nonumber \\
P_s&\approx & \frac{n n!}{12(1+n)^n}\beta \left[6+(6+2n+n^2)\beta \right],
\label{appTh}
\end{eqnarray}
where $t$ substitutes the squeezing by $t=1-V$.  Those expressions parametrize the thresholds tightly even beyond the considered regime of weak states and they can be use as a better approximate. Their reliability was checked by Monte - Carlo simulation up to order five. The results of simulation related to the fourth and fifth ordered criterion are illustrated in Fig.~\ref{MCsimSM}. Assuming $t\ll 1$, the criteria get simpler forms
\begin{equation}
 P_e<\frac{(1+n)^{2n}(2+n)^2(1+n)!P_s^3}{18n^2(n!)^3},
 \label{appendixTh}
 \end{equation}
which however diverge quickly from the exact threshold when $P_e$ grows.

\begin{figure}

\includegraphics[width=\columnwidth]{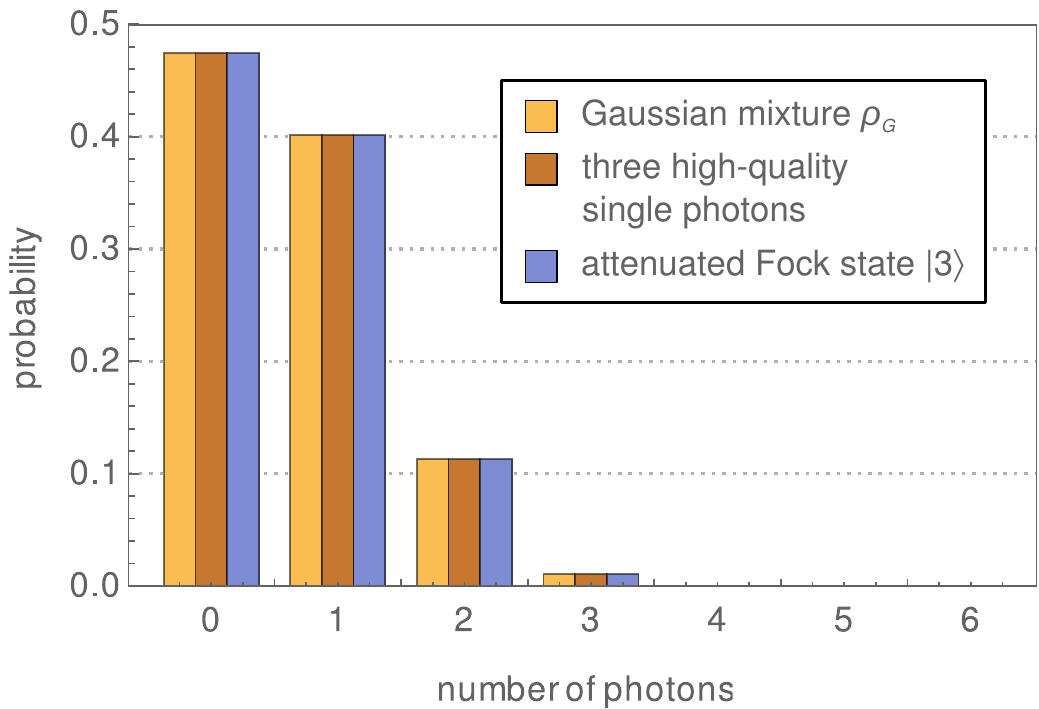}

\vspace{12pt}

\includegraphics[width=\columnwidth]{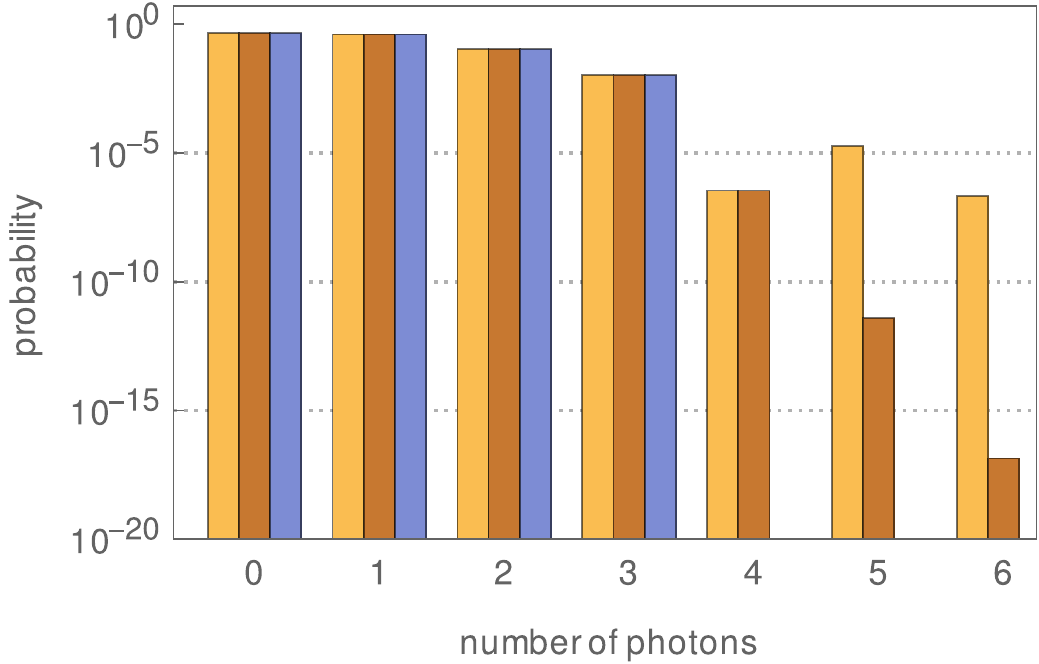}

\vspace{12pt}

\includegraphics[width=\columnwidth]{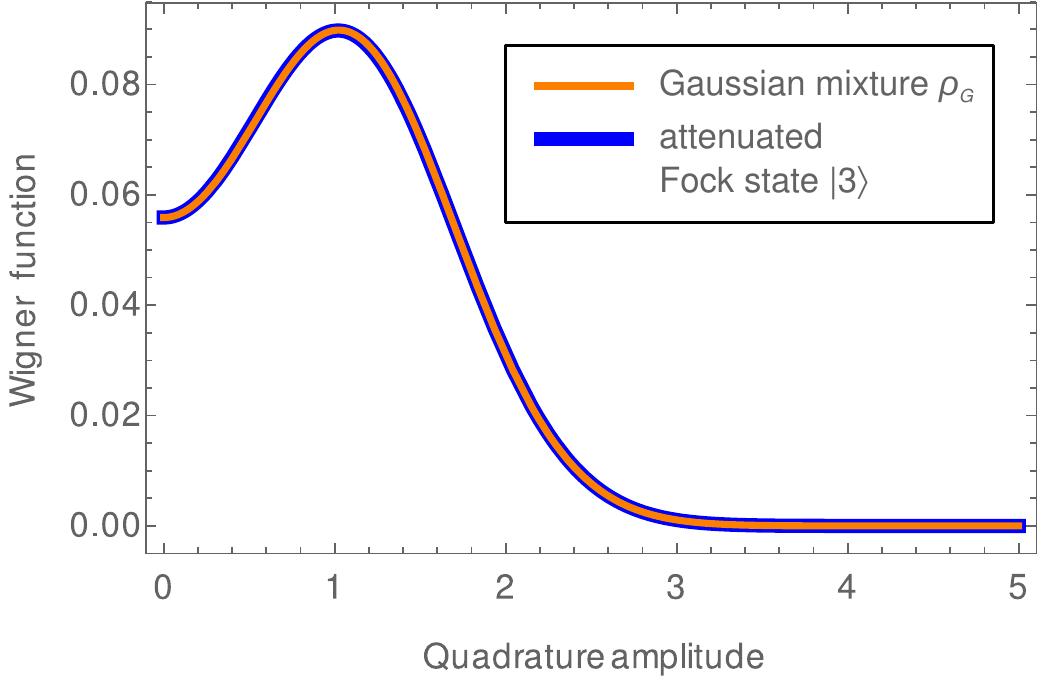}

\caption{Comparison of a Gaussian mixture and a genuine quantum non-Gaussian states presented in the text. The statistical distribution of photons is depicted in linear (top sub-figure) and \emph{log-log} scale (middle sub-figure). The radial profiles of the Wigner functions of all the states are plotted in the bottom sub-figure. The difference among them is in the order of $10^{-4}$.}
\label{fig.comparison}
\end{figure}

The analysis can be demonstrated in a fine distinction of $3$-photon genuine quantum non-Gaussianity among examples of three different states of light. Let us consider an ideal Fock state $\vert 3 \rangle$ affected by losses $T=0.22$. This state manifests from definition the genuine quantum non-Gaussianity of order $n=1,2,3$ because it exhibits perfect truncation of photon statistics. A next considered state is represented by three copies of a high quality single photon state having photo-distribution $p_1=0.22$, $p_2=2.4 \times 10^{-6}$ and $p_0=1-p_1-p_2$. The realistic contributions of three photons can be neglected in this example. After an analysis, the state also manifests $3$-photon genuine quantum non-Gaussianity. And finally, the last considered state is a statistical mixture of Fock states $\vert 0 \rangle$, $\vert 1 \rangle$ and $\vert 2 \rangle$ modified by squeezing and displacement operations. We define it by taking $\vert \phi \rangle = S(\beta)D(\alpha)\vert 2 \rangle$ with $\alpha=0.16747 e^{i\phi}$ and $\beta=0.014044 e^{i \phi}$, randomizing its phase and mixing it with lower Fock states
\begin{eqnarray}
\rho_G &=& 0.47425 \times \vert 0 \rangle \langle 0 \vert +0.39519 \times \vert 1 \rangle \langle 1 \vert+ \nonumber \\
&\ &+0.13056 \times \int_0^{2\pi} \vert \phi \rangle \langle \phi \vert \mathrm{d}\phi.
\end{eqnarray}
This state does not exhibit $3$-photon genuine quantum non-Gaussianity from definition. 

A comparison of these states is illustrated in Fig.~7. The statistical properties are depicted in the top and middle sub-figure. The dominant contributions of all three states are almost identical. A main difference appears for probabilities of four and more photons as can be seen in the middle sub-figure with \emph{log-log} scale. Also the radial profile of the Wigner function in the bottom sub-figure indicates the states behave very similarly. This inspection demonstrates that although the Gaussian mixtures which do not exhibit $3$-photon genuine quantum non-Gaussianity can be very similar to the states which do exhibit this property, the difference among the almost suppressed events is crucial for the diagnostics.

The experimental demonstration of genuine $n$-photon quantum non-Gaussianity  is completed by information about robustness against losses and additional noise, which can be theoretically predicted in the region of states with low probability of error. In those cases, a click statistics is contributed approximately only by
\begin{eqnarray}
P_s & \approx & \frac{n!}{(n+1)^n} \rho_n \nonumber \\
P_e & \approx & \frac{n!}{(n+1)^n} \rho_{n+1},
\label{statisticsApp}
\end{eqnarray}
where $\rho_k$ is a probability a state $\rho$ has $k$ photons, i. e. $\rho_k=\langle k \vert \rho \vert k \rangle$. Transmission efficiency $T$ changes the photon statistics approximately so this $\rho_k \rightarrow \rho_k T^k$.
In the $log-log$ scale, a measured point follows a line
\begin{equation}
\log_{10} P_e^{(T)} = \frac{n+1}{n} (\log_{10} P_s^{(T)}-\log_{10} P_s)+ \log_{10} P_e,
\end{equation}
where $P_{s,e}^{(T)}$ are measured probabilities affected by attenuation and $P_{s,e}$ are the original ones. Employing this rule together with (\ref{appendixTh}) enable to estimate the robustness against losses. Poissonian noise influence the statistics by
\begin{eqnarray}
P_s & \approx & \frac{n!}{(n+1)^n} (\rho_n + \bar{n} \rho_{n-1}) \nonumber \\
P_e & \approx & \frac{n!}{(n+1)^n} (\rho_{n+1}+ \bar{n} \rho_{n}),
\label{noiseApp}
\end{eqnarray}
where $\bar{n}$ is a mean number of photons of the noise. Because the criteria impose a very strict condition on the noise, one can assume the members contributing the success probability satisfy $\bar{n} \rho_{n-1} \ll \rho_n$. Therefore a point moves in the plots vertically along the $\log_{10} P_e$ axis. However, the data show slight dropping of probability $P_s$ for states affected by the background noise, as apparent in the Fig.~2 of the main part of the manuscript. It arises from imperfect protection of the heralding detector from photons of the noise in the experiment. Thus, a heralding event was rarely caused by the background noise that decrease slightly probability $P_s$.

\flushbottom

\end{document}